\documentclass[12pt]{article}
\usepackage{a4wide,epsfig,amsmath,amssymb,cite,scalefnt,graphicx}

\def\npb{{\it{Nucl.\ Phys.}\ }{\bf B}}
\def\plb{{{\it Phys.\ Lett.}\ }{ \bf B}}
\def\jhep{{\it JHEP}\ }
\def\be{\begin{equation}}
\def\ee{\end{equation}}
\def\bea{\begin{eqnarray}}
\def\eea{\end{eqnarray}}
\def\nn{\nonumber\\}

\def\tr{\hbox{tr}}

\def\prd{{{\it Phys.\ Rev.}\ }{\bf D}}

\def\thetabar{{\overline\theta}}

\def\cbar{\overline c}
\def\fbar{\overline f}
\def\Fbar{\overline F}
\def\Lambdabar{\overline \Lambda}

\def\Tr{\mathop{\rm Tr}}

\def\frakk[#1#2{{{#1}\over{#2}}}

\def\lambdabar{\overline\lambda}

\def\Dbar{\overline D}

\def\Phibar{\overline\Phi}

\def\Fbar{\overline F}

\def\Ncal{{\cal N}}

\def\pa{\partial}
\def\pabar{\overline\partial}

\input epsf
\begin{document}

\begin{titlepage}
\begin{flushright}
LTH975\\
\end{flushright}
\date{}
\vspace*{3mm}

\begin{center}
{\Huge Four-loop results for the anomalous dimension 
for a general $N=2$ supersymmetric 
Chern-Simons theory in three dimensions}\\[12mm]
{\bf I.~Jack\footnote{{\tt dij@liv.ac.uk}} and 
C.~Luckhurst\footnote{{\tt mf0u60d7@liv.ac.uk}}}\\

\vspace{5mm}
Dept. of Mathematical Sciences,
University of Liverpool, Liverpool L69 3BX, UK\\

\end{center}

\vspace{3mm}
\begin{abstract}
We present results for the planar contribution to the 
four-loop anomalous dimension for a general $N=2$ supersymmetric 
Chern-Simons theory in three dimensions. These results should facilitate
higher-order superconformality checks for theories relevant for the AdS/CFT 
correspondence. 

\end{abstract}

\vfill

\end{titlepage}

\section{Introduction}
Chern-Simons gauge theories have attracted attention for a considerable 
time due to their topological nature\cite{schwarza, witten, djt}
(in the pure gauge case) and their 
possible relation to the quantum Hall effect and high-$T_c$ 
superconductivity. More recently  
there has been substantial interest in $\Ncal=2$ supersymmetric Chern-Simons
matter theories in the context of the AdS/CFT 
correspondence and in particular, a wide range of superconformal theories has 
been discovered\cite{schwarz}-\cite{MFMa}, starting with the 
BLG\cite{bagger,gust} and ABJ/ABJM\cite{ABJ,ABJM} models.
Although a more familiar formulation is in terms of ``quiver''-type gauge 
theories, many of them may be understood in terms
of an underlying ``3-algebra'' 
structure\cite{bagger},\cite{baggera}-\cite{CWf}. 
Explicit perturbative computations to
corroborate the superconformal property have been carried out in 
Refs.~\cite{ASW, penatia, penatib}
at lowest order (two loops for a theory in three dimensions). Since
the gauge coupling is unrenormalised for any Chern-Simons theory due to
the topological nature of the theory, it is only necessary to compute 
the anomalous dimensions of the chiral fields in order to check for 
superconformality (in view of the non-renormalisation theorem). Our purpose here
is to provide results to enable the extension of this check to the next 
(four-loop) order. As may readily be imagined, this is a highly non-trivial   
undertaking. Consequently we envisage an abridged version of the full task.
Firstly, we have calculated only the planar diagrams, corresponding to the
leading $N$ contributions. Even then, and even after 
discarding large classes of diagrams which can be seen in advance not
to contribute to the anomalous dimension, one is faced with of the order
of a hundred distinct 
diagrams. The process of automation which has made it feasible to
perform high-loop calculations in non-supersymmetric theories using 
packages such as {\sc Mincer} is not available to us here; we are not aware
of any available package for performing superspace calculations.  
Secondly, therefore, we have confined ourselves to computing
the coefficients of only a subset (albeit a large one) of the invariants
contributing to the anomalous dimension. We believe that the remaining
coefficients may be established by {\it assuming} the superconformality
of a small number of the known examples of such 
theories. Armed with the full planar result, one could then continue
to check the remaining ones, together with any new examples proposed in future.
To put it another way, the conditions required for four-loop
superconformality of all the known examples of such theories will form a 
highly redundant set of consistency conditions. Our lack of knowledge 
of all the coefficients in the anomalous dimension at this order will reduce
this redundancy somewhat but there should be enough remaining to give 
considerable confidence in the persistence of the superconformality property
to this order. Moreover we have tried to facilitate an extension of the check
to include further invariants in the anomalous dimension, in the following 
sense: for the subset of invariants on which we have focussed our
attention, we have (of course) computed all the diagrams which can contribute.
Many of these diagrams also contribute to other invariants, and in these 
cases we have listed the contributions to these other invariants so that they 
can readily be combined with the contributions from the remaining diagrams
at a later date.   

The paper is organised as follows: in Section 2 we describe the general 
$\Ncal=2$ supersymmetric Chern-Simons theory in three dimensions; in Section 3 
we describe our calculations and give our result; Section 4 is devoteed to a 
discussion of the result and suggestions for future work; while we explain
our conventions and list various useful basic results and identities in an 
Appendix.

\section{$\Ncal=2$ Chern-Simons theory in three dimensions}
The action for the theory can be written 
\be
S=S_{SUSY}+S_{GF}
\label{lag}
\ee
where $S_{SUSY}$ is the usual supersymmetric action\cite{ivanov}
\bea
S_{SUSY}=&\int d^3x\int d^4\theta\left(
k\int_0^1dt\Tr[\Dbar^{\alpha}(
e^{-tV}D_{\alpha}e^{tV})]+\Phi^j (e^{V_AR_A})^i{}_j\Phi_i\right)\nn
&+\left(\int d^3x\int d^2\theta W(\Phi)
+\hbox{h.c.}\right).
\label{ssusy}
\eea
Here $V$ is the vector
superfield, $\Phi$ the chiral matter superfield and the 
superpotential (quartic for renormalisability in three dimensions)
$W(\Phi)$ is given by
\be
W(\Phi)=\frac{1}{4!}Y^{ijkl}\Phi_i\Phi_j\Phi_k\Phi_l.
\ee
(We use the 
convention that $\Phi^i=(\Phi_i)^*$.) 
We assume a simple gauge group, though we comment later on the extension to
non-simple groups.
Gauge invariance requires the gauge coupling $k$ to be 
quantised, so that $2\pi k$ is an integer. The vector superfield $V$ is in the
adjoint representation, $V=V_AT_A$ where $T_A$ are the generators of the
fundamental representation, satisfying 
\bea
[T_A,T_B]=&if_{ABC}T_C,\nn
\Tr(T_AT_B)=&\delta_{AB}.
\eea
The chiral superfield can be in a general representation, with gauge
matrices denoted $R_A$ satisfying
\bea
[R_A,R_B]=&if_{ABC}R_C,\nn
\Tr(R_AR_B)=&T_R\delta_{AB}.
\label{Tdef}
\eea
In three dimensions the Yukawa couplings $Y^{ijkl}$ are dimensionless 
and (as mentioned earlier) the theory is renormalisable. 
In Eq.~(\ref{lag}) the gauge-fixing term $S_{GF}$ is given by\cite{penatib} 
\be
S_{GF}=\frac{k}{2\alpha}\int d^3xd^2\theta\tr[f\fbar]
-\frac{k}{2\alpha}\int d^3xd^2\thetabar\tr[f\fbar]
\label{sgf}
\ee 
and we introduce into the functional integral a corresponding ghost term
\be  
\int{\cal D}f{\cal D}\fbar\Delta(V)\Delta^{-1}V
\ee
with 
\be
\Delta(V)=\int d\Lambda d \Lambdabar\delta(F(V,\Lambda,\Lambdabar)-f)
\delta(\Fbar(V,\Lambda,\Lambdabar)-\fbar),
\label{ghostdet}
\ee
with $\Fbar=D^2V$, $F=\Dbar^2V$.
With $\alpha=0$ this results in a gauge propagator
\be
\langle V^A(1)V^B(2)\rangle=-\frac1K\frac{1}{\pa^2}
\Dbar^{\alpha}D_{\alpha}\delta^4(\theta_1-\theta_2)\delta^{AB}.
\ee
The gauge vertices are obtained by expanding $S_{SUSY}+S_{GF}$ as given by
Eqs.~(\ref{ssusy}), (\ref{sgf}): 
\bea
S_{SUSY}+S_{GF}&\rightarrow&
-\frac{i}{6}f^{ABC}\int d^3xd^4\theta\Dbar^{\alpha}V^A
D_{\alpha}V^BV^C\nn
&-&\frac{1}{24}f^{ABE}f^{CDE}\int d^3xd^4\theta\Dbar^{\alpha}V^AV^B
D_{\alpha}V^CV^D+\ldots.
\eea

The ghost action resulting from Eq.~(\ref{ghostdet}) has the same form
as in the four-dimensional $\Ncal=1$ case\cite{GGRS,GRS}
\be
S_{gh}=\int d^3xd^4\theta\tr\{\cbar'c-c'\cbar+
\frac12(c+\cbar')[V,c+\cbar)]+
\frac{1}{12}(c+\cbar')[V,[V,c-\cbar)]]+\ldots\}
\label{sgh}
\ee
leading to ghost propagators
\be
\langle \cbar'(1)c(2)\rangle=-\langle c'(1)\cbar(2)\rangle
=-\frac{1}{\pa^2}\delta^4(\theta_1-\theta_2).
\ee
and cubic, quartic vertices which may easily be read off from Eq.~(\ref{sgh}).
Finally the chiral propagator and chiral-gauge vertices are readily obtained
by expanding Eq.(\ref{ssusy}); the chiral propagator is given by:
\be
\langle \Phi^i(1)\Phi_j(2)\rangle=-\frac{1}{\pa^2}\delta^4(\theta_1-\theta_2)
\delta^i{}_j.
\ee
The regularisation of the theory is effected by replacing 
$V$, $\Phi$, $Y$ by corresponding bare quantities $V_B$, $\Phi_B$, $Y_B$,
with the bare and renormalised fields related by  
\be
V_B=Z_VV,\quad \Phi_B=Z_{\Phi}\Phi.
\ee 
Since the Chern-Simons level $k$ is expected to be unrenormalised for a 
generic Chern-Simons theory due to the topological nature of the theory
(so that $k_B=k$), 
superconformality will be determined purely by the vanishing of the 
$\beta$-functions for the superpotential coupling. These will be 
given according to the non-renormalisation theorem by
\be
\beta_Y^{ijkl}=\gamma_{\Phi m}^{(i}Y^{jkl)m}.
\label{nonren}
\ee 
where the anomalous dimension $\gamma_{\Phi}$ is defined by 
\be
\gamma_{\Phi}=\mu\frac{d}{d\mu}\ln Z_{\Phi}.
\ee
Writing
\be
Z_{\Phi}=\sum_{L\,\rm{even},m=1\ldots \frac{L}{2}}
\frac{Z_{\Phi}^{(L,m)}}{\epsilon^m}
\ee
$\gamma_{\Phi}$ is determined by the simple poles in $Z_{\Phi}$ according
to
\be
\gamma_{\Phi}^{(L)}=LZ_{\Phi}^{(L,1)}
\ee
and the 
higher order poles in $Z_{\Phi}$ 
are determined by consistency conditions, the one
relevant for our purposes being 
\be
Z_{\Phi}^{(4,2)}=\beta_Y^{(2)}.\frac{\pa}{\pa Y}\gamma_{\Phi}^{(2)}
-2\left(\gamma_{\Phi}^{(2)}\right)^2
\label{doublep}
\ee
where $\beta_Y$ is given by Eq.~(\ref{nonren}) and
\be
\beta_Y.\frac{\pa}{\pa Y}\equiv
\beta_Y^{klmn}.\frac{\pa}{\pa Y^{klmn}}.
\ee
At lowest order (two loops) it was found that superconformality (i.e. 
the vanishing of $\beta_Y$) was equivalent to the vanishing of $\gamma_{\Phi}$
in all the cases considered\cite{penatib, ASW} and it appears likely that this
will remain true at higher orders.

\section{Perturbative Calculations}
In this section we review the two-loop calculation and describe in 
detail our four-loop results.

The anomalous dimension of the chiral superfield is given at two loops
by\cite{penatib,ASW}
\be
(8\pi)^2\gamma_{\Phi}^{(2)}=
\frac13Y_2-2k^{-2}C_RC_R-k^{-2}T_RC_R+k^{-2}C_GC_R
\label{gamphi}
\ee  
where
\bea
(Y_2)^i{}_j=&Y^{iklm}Y_{jklm}\nn
C_R=&R_AR_A,\nn
C_G\delta_{AB}=&f_{ACD}f_{BCD}
\eea
and $T_R$ is defined in Eq.~(\ref{Tdef}).
This result may readily be obtained by $\Ncal=2$ superfield 
methods\cite{agk,GN,ASW,penatib}; see
the Appendix for our $\Ncal=2$ superfield conventions. Henceforth we set
$k=1$ for simplicity; it may easily be restored if desired. Two-loop
results for general Chern-Simons theories have also been obtained in 
Ref.~\cite{akk} but are not directly comparable since they were computed
in the $N=1$ framework. 

\begin{figure}
\includegraphics{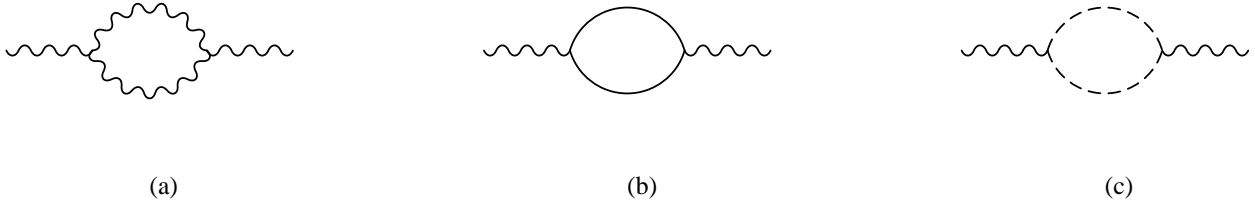}
\caption{The one-loop insertions contributing to $X$}
\label{Xdiag}
\end{figure}

As explained earlier, in this paper we confine ourselves to the 
contributions to the four-loop anomalous dimension from planar 
diagrams. From a consideration of possible group invariants, the four-loop 
anomalous dimension is expected to take the form
\bea
(8\pi)^4\gamma_{\Phi}^{(4)}&=&
\alpha_1 Z_1+\alpha_2 Z_2 +\alpha_3W_1+\alpha_4W_2+\alpha_5W_3
+\alpha_6W_4+(\alpha_7 X+\alpha_8C_G)U_1\nn
&+&(\alpha_9 X+\alpha_{10}C_G)U_2
+\alpha_{11}C_{40}+\alpha_{12}C_{31}+\alpha_{13}C_{22}+\alpha_{14}C_{13}+
\alpha_{15}F_4\nn
&+&X\left(\alpha_{16}C_{30}+\alpha_{17} C_{21}+\alpha_{18} C_{12}\right)
+X^2(\alpha_{19}C_{20}+\alpha_{20}C_{11})+\alpha_{21}X^3C_R\nn
&+&\alpha_{22}X_2
+\alpha_{23}X_4
+(\alpha_{24}X+\alpha_{25}C_R
+\alpha_{26}C_G)X_1\nn
&+&\alpha_{27}X_5C_R
+(\alpha_{28}X+\alpha_{29}C_G)X_5+\alpha_{30}X_3\nn
&+&\alpha_{31}\tr(C_R\{R_A,R_B\}R_C)R_AR_BR_C
+\alpha_{32}X_6+\alpha_{33}d_{CDA}d_{CDB}R_AR_B
\eea
where the invariants involving Yukawa couplings are given by
\bea
(Y_3)^i{}_j&=&Y^{ikmn}(Y_2)^l{}_kY_{jlmn},\nn
(Y_4)^i{}_j&=&Y^{iklr}Y_{klmn}Y^{mnpq}Y_{pqrj},\nn
Z_1&=&Y_2C_RC_R,\nn
(Z_2)^i{}_j&=&Y^{iklm}Y_{jkln}(C_RC_R)^n{}_m\nn
(W_1)^i{}_j&=&Y^{iklm}Y_{rknp}(R_A)^n{}_l(R_B)^p{}_m(R_AR_B)^r{}_j,\nn
(W_2)^i{}_j&=&Y^{iklm}Y_{pkln}(R_AR_B)^n{}_m(R_BR_A)^p{}_j,\nn
(W_3)^i{}_j&=&Y^{ikmp}Y_{jkln}(R_AR_B)^l{}_m(R_AR_B)^n{}_p,\nn
(W_4)^i{}_j&=&Y^{iklm}Y_{pkln}(R_AC_R)^n{}_m(R_A)^p{}_j,\nn
U_1&=&Y_2C_R,\nn
(U_2)^i{}_j&=&Y^{iklm}Y_{jkln}(C_R)^m{}_n,
\label{Ystruct}
\eea
and the remaining ones are
\bea
C_{mn}&=&C_R^mC_G^n,\nn
F_4&=&f_{EAB}f_{ECD}f_{HAF}f_{HCG}R^BR^DR^FR^G,\nn
X&=&T_R-\frac12C_G\nn
X_1&=&\tr(C_RR_AR_B)R_AR_B,\nn
X_2&=&\tr(R_AR_BR_CR_D)R_AR_BR_CR_D,\nn
X_3&=&\tr(C_RC_RR_AR_B)R_AR_B,\nn
X_4&=&\tr(Y_2R_AR_B)R_AR_B,\nn
X_5&=&D_{ABC}R_AR_BR_C,\nn
X_6&=&f_{EAB}D_{ECD}R_AR_CR_BR_D,
\label{gstruct}
\eea
with
\be
D_{ABC}=\frac12\tr(\{R_A,R_B\}R_C).
\ee
The quantity $X$ in Eq.~(\ref{gstruct}) is produced by one-loop vector 
two-point insertions as depicted in Fig.~\ref{Xdiag}. 
One can show using results from Ref.~\cite{macfar} that the structure $X_6$ 
vanishes for the case of the fundamental representation; but we have not been
able to prove this in general. We have decided to omit the computation of the 
coefficients $\alpha_{14,15}$, $\alpha_{18}$, $\alpha_{20}$, $\alpha_{26}$,
$\alpha_{29}$ and $\alpha_{30-33}$, and therefore we shall leave out those 
diagrams which can only contribute to these coefficients. Our rationale broadly 
speaking has been to avoid coefficients which derive contributions from large 
numbers of diagrams. This typically entails avoiding invariants with factors 
of $C_G$, since it is clear for instance from Table~\ref{group} that 
invariants with more factors of $C_G$ can arise from a larger number of 
diagrams. The coefficients $\alpha_{12,13}$, $\alpha_{17}$ are
exceptions to this; we computed these since the corresponding invariants
$C_{31}$, $C_{22}$ and $XC_{21}$ have non-zero
double poles (see Eq.~(\ref{doubexp})), which we wished to compute as a 
consistency check.  

\begin{figure}
\includegraphics{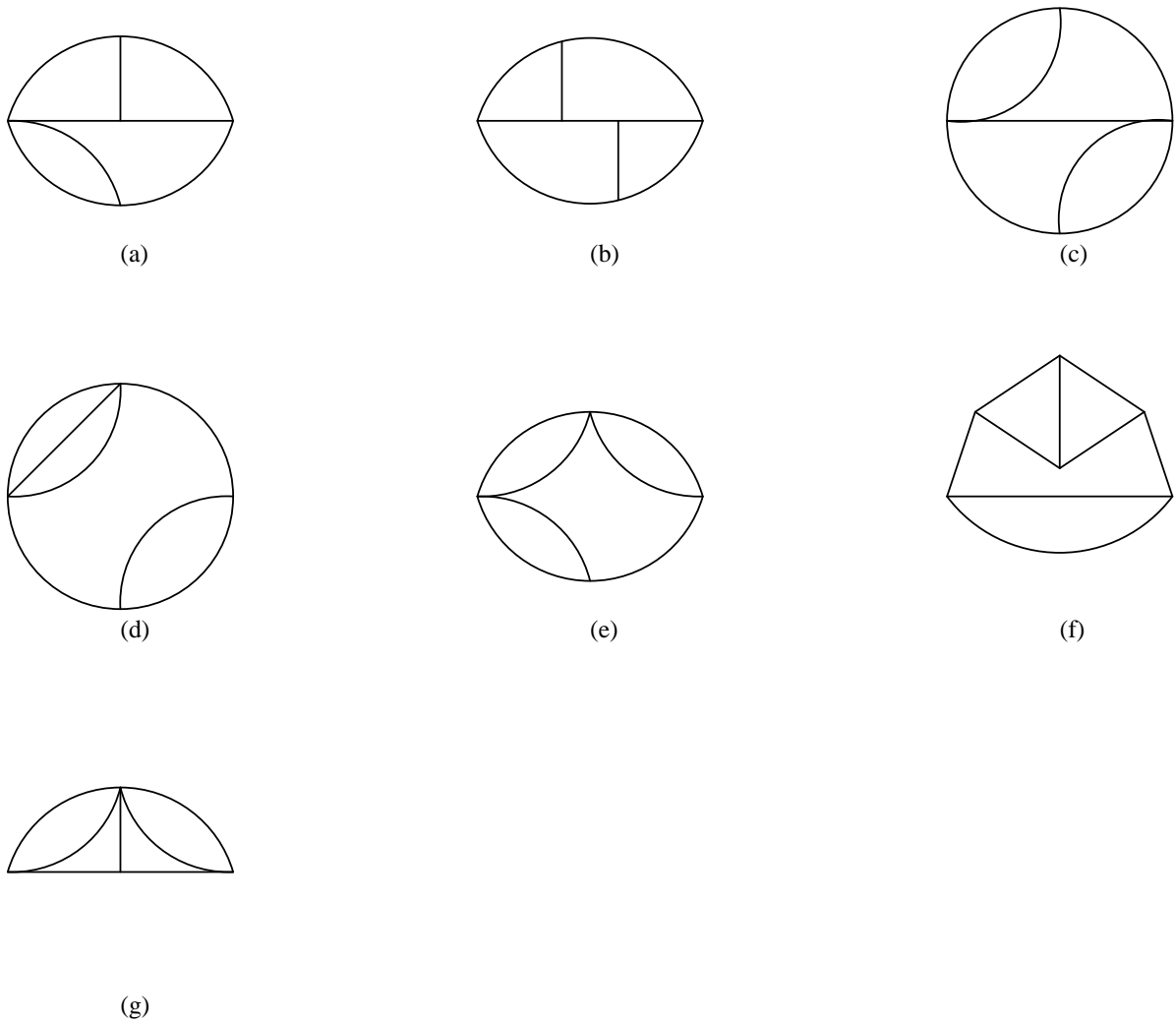}
\caption{Four-loop topologies}
\label{tops}
\end{figure}

We are therefore
concerned with the calculation of two-point diagrams.
We have used the package {\it FeynArts}\cite{hahn} 
to assist in generating the full 
set of diagrams.
This package requires as an input the basic four-loop planar vacuum topologies,
since only the topologies up to three loops are contained in the 
standard package. The topologies which we have used in {\it FeynArts} are depicted in 
Fig.~\ref{tops} and also in Fig.~\ref{momfour}; the remaining topologies 
consist of insertions of loops on simple three-loop topologies and we have 
enumerated diagrams in these classes 
``by hand''. We can provide the topology files upon request to anyone
interested in checking or extending our calculations.

\begin{figure}
\includegraphics{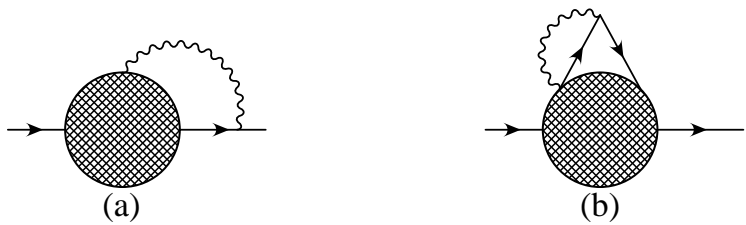}
\caption{Classes of diagram which do not contribute}
\label{fig2}
\end{figure}

Two large classes
of diagrams may be immediately discarded as having no logarithmic 
divergences and therefore no contribution to the anomalous dimension
\cite{GN}. The first
consists of those diagrams in which the first (last) vertex encountered along 
the incoming (outgoing) chiral line has a single gauge line. These are shown
schematically in Fig.~\ref{fig2}(a).
The second class consists of those diagrams which
contain a one-loop subdiagram with one gauge and one chiral line; depicted
in Fig.~\ref{fig2}(b).
We were able to use the features of {\it FeynArts} to discard such
diagrams of the type in Fig.~\ref{fig2}(a) automatically.

\begin{figure}
\includegraphics{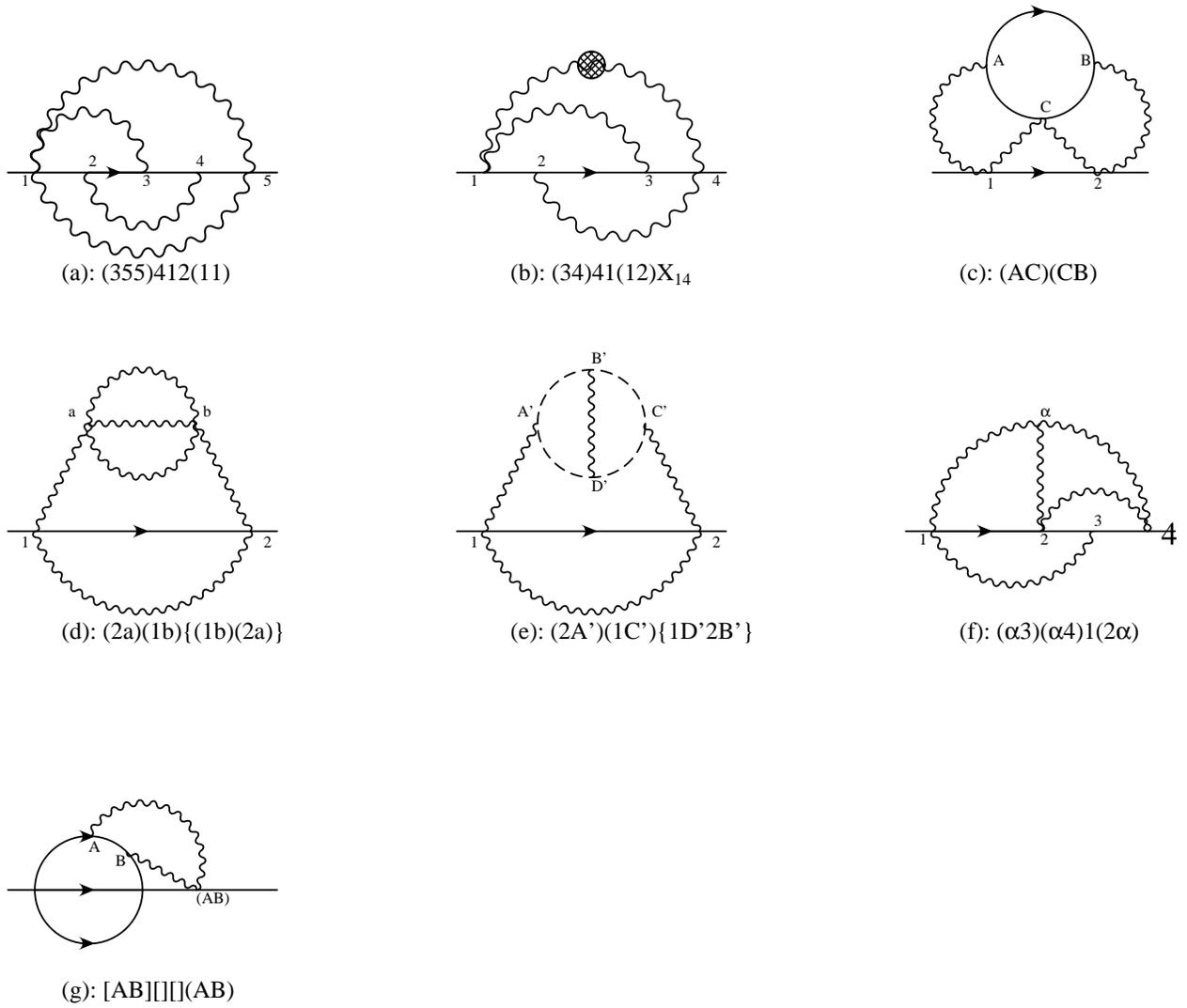}
\caption{Typical diagrams with their notation}
\label{fig3}
\end{figure}

Instead of displaying each of the divergent graphs pictorially, 
which would be very laborious,
we introduce a notation for various classes of diagram and illustrate it by 
means of representative examples, depicted in Fig.~\ref{fig3}. 
The main exception is the graphs with Yukawa 
vertices, which we shall describe shortly.
The majority of our graphs have no 
Yukawa vertices and most can be described using a fairly uniform notation.
We start with graphs which have only a single chiral line, and only 
gauge/chiral vertices. 
The gauge-chiral vertices along the chiral line are labelled numerically in 
order from left to right. The structure of the diagram is then indicated by 
recording to which other vertices each vertex is connected, again from left
to right. An example is depicted in Fig.~\ref{fig3}(a).

\begin{figure}
\includegraphics{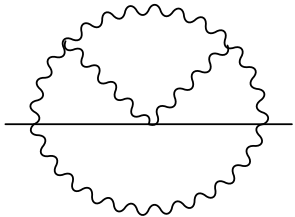}
\caption{The additional graph from Table~\ref{restabc}}
\label{fig3a}
\end{figure}

The insertion of a one-loop gauge 2-point function on the propagator joining
vertices $i$ and $j$ is denoted by the addition of $X_{ij}$; see 
Fig.~\ref{fig3}(b). 
More complex chiral loops (but without internal lines) are described by 
labelling the vertices on the loop 
by $A$, $B$, alphabetically, following the direction of the chiral arrows, and 
listing their connections to the vertices on the ``main'' chiral line as in the 
previous examples; as in Fig.~\ref{fig3}(c). 
Gauge loops are described similarly, but with 
lower-case letters. If there are internal lines within the loop, these
are denoted by listing the connections alphabetically in the same way as the 
main chiral line, but enclosed within brackets $\{\}$, as in 
Fig.~\ref{fig3}(d). Ghost 
loops are denoted by labelling their vertices with primed letters, as in 
Fig.~\ref{fig3}(e). 
Gauge vertices which do not lie within loops are denoted by greek
letters and their connections with the main line or with loops denoted as usual,
as in Fig.~\ref{fig3}(f). A single graph which does not fall into any of these 
categories is that shown 
in Fig.~\ref{fig3a} (the result for this graph will be given in 
Table~\ref{restabc}).

\begin{figure}
\includegraphics{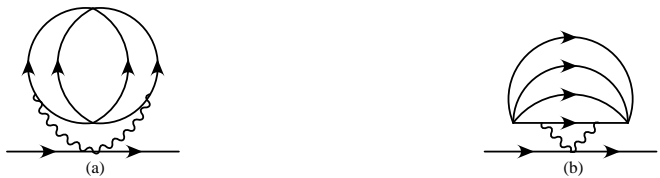}
\caption{The graphs of Table~\ref{restabg}}
\label{fig4}
\end{figure}

\begin{figure}
\includegraphics{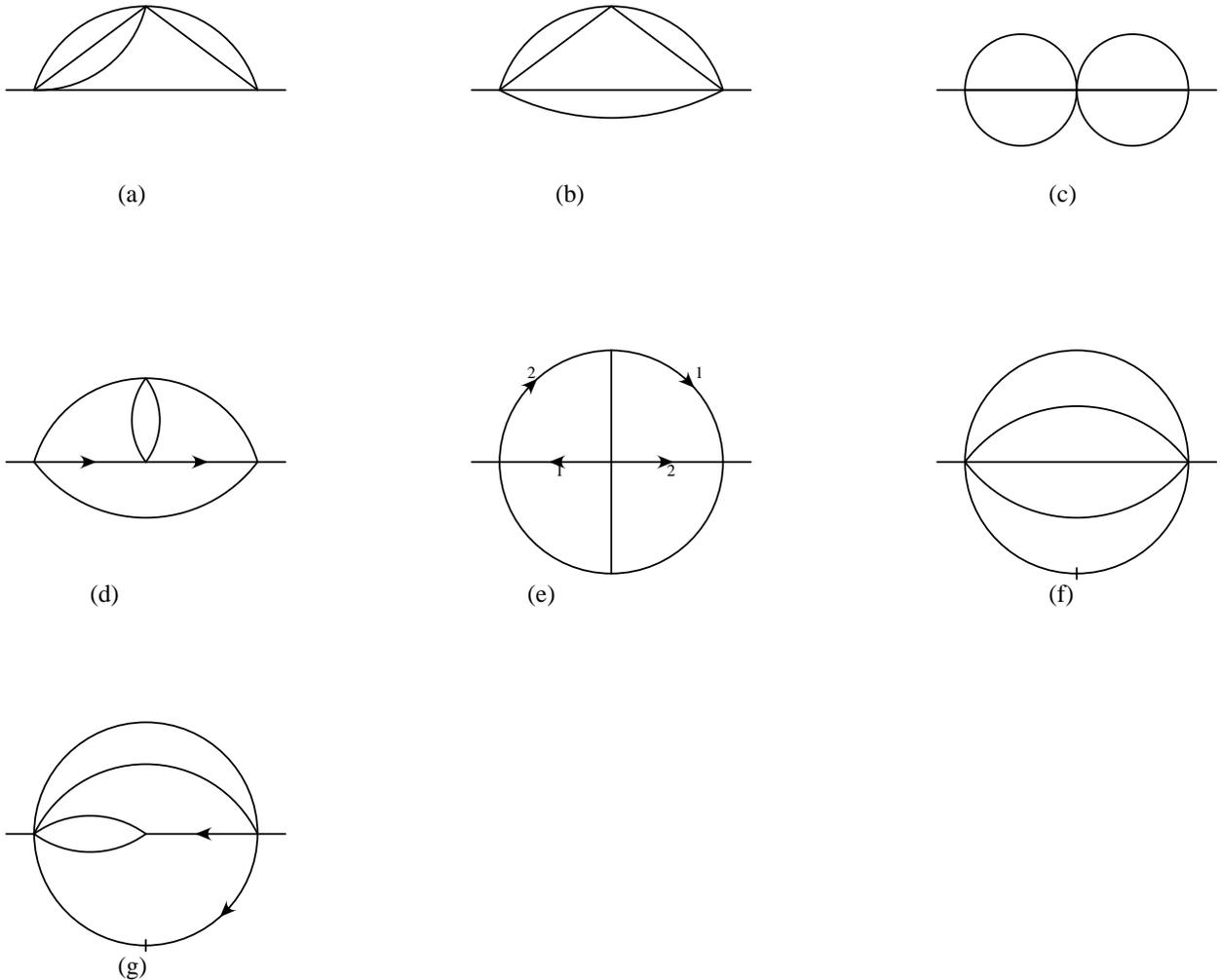}
\caption{The basis of momentum integrals}
\label{momfour}
\end{figure}

Now we come to the graphs with Yukawa vertices. There are two graphs with
four Yukawa vertices (and no gauge lines). Their structure can easily 
be derived from the corresponding group invariant (one for each graph). They 
will therefore simply be labelled by their
group structure ($Y_3$ and $Y_4$, as defined in Eq.~(\ref{Ystruct})).
The graphs with two Yukawa vertices (connected to an external line) and two
gauge lines are described using a somewhat different 
notation to the above. 
The gauge matrices on chiral lines are labelled $A$, $B$, etc and the matrices
on each chiral line in the chiral loop are enclosed within square brackets
$[]$. Two matrices labelled with the same letter are connected by a gauge 
propagator. This is exemplified in Fig.~\ref{fig3}(g). Finally, 
we have found it simplest to depict the diagram explicitly for a 
small class of diagrams with two Yukawa vertices, in Fig.~\ref{fig4}.

Several diagrams clearly give no contribution by virtue of group theoretic 
considerations. For each remaining diagram the $D$-algebra is performed using 
the conventions and useful 
identities listed in the Appendix. A large number (almost all, in fact) of 
diagrams
containing 3-point gauge vertices yield vanishing contributions when the 
results of all possible arrangements of the $D$s and $\Dbar$s are added
together. Unfortunately we have not succeeded in establishing a criterion
to predict in advance which diagrams give non-vanishing results. 
Our results for the non-vanishing divergences are listed diagram-by-diagram in 
Tables~\ref{restaba}-\ref{restabl}. 
Note that the graph $(3\alpha)\alpha\alpha(1\alpha)$ in 
Table~\ref{restabc} yields two distinct group structures which have been listed
separately, the second occurrence distinguished by a prime.

Let us now explain how the Tables have been constructed.
The results have been expressed in terms
of a relatively small basis of momentum integrals\cite{MSS,MSSa}
which are depicted in Fig.~\ref{momfour}
and whose divergences are also listed in the 
Appendix. Figs.~\ref{momfour}(a)-(g) depict $I_4$, $I_{4bbb}$, $I_{22}$,
$I_{42bbc}$, $I_{422qAbBd}$, $J_4$ and $J_5$, respectively. The results 
given later, and also most of these conventions for labelling the diagrams, 
are taken from Refs.~\cite{MSS,MSSa}. 
In Fig.~\ref{momfour} the arrows denote momenta in the numerator
contracted as indicated. 
These momentum integrals multiply a variety of group structures,
which appear in the final columns of Tables~\ref{restaba}-\ref{restabl}.
In Table~\ref{group} we give the decompositions of some of these group 
structures into the basis of group invariants. The definitions of these group 
structures are not given explicitly as they may easily be read off from
the structure of the diagrams where they appear. Two examples should 
suffice: for instance, to take the diagrams Fig.~\ref{fig3}(a), (b) 
respectively
\bea
S_4&=&R_{(A}R_BR_{C)}R_DR_AR_DR_{(B}R_{C)}\nn
S_{X4}&=&XR_{(A}R_{B)}R_CR_AR_{(B}R_{C)}.
\eea
Finally the first columns of Tables~\ref{restaba}-\ref{restabl} contain
an overall symmetry factor. The resulting contribution to the 
two-point function for each diagram is therefore obtained 
by adding the momentum integrals with the coefficients listed in the appropriate
row and multiplying the resulting sum by the corresponding symmetry factor and 
group structure. For instance, the fourth row of Table~\ref{restaba} denotes
a contribution
\be
(-1)(-2I_4+I_{4bbb})\left(W_2-\frac{1}{12}C_GU\right).
\ee
The combination of momentum integrals 
\be
\frac14I_4-\frac58I_{22}-I_{4bbb}+I_{42bbc}-2I_{422qAbBd},
\ee
which one frequently observes in the tables, results from a momentum integral
corresponding to the topology Fig.~\ref{momfour}(e), but with a trace over 
a product of ``$p_{\alpha\beta}$'' around the perimeter (constructed as in
Eq.~(\ref{pdef}), where $p$ is the 
momentum on one of the perimeter lines).

\begin{figure}
\includegraphics{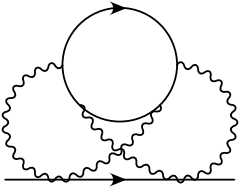}
\caption{The non-planar graph}
\label{nonplanar}
\end{figure}

The first check on our results is provided by the consistency 
conditions Eq.~(\ref{doublep}) for the double poles. These, with the aid
of Eq.~(\ref{gamphi}), give
\bea
(8\pi)^4Z_{\Phi}^{(4,2)}&=&\frac16Y_3
-\frac{1}{12}Z_1+\frac14Z_2+\frac18\left(U_2-\frac13U_1\right)
\left(X-\frac12C_G\right)\nn
&+&\frac12C_{40}-\frac14C_{31}+\frac{1}{32}C_{22}+\frac12XC_{30}
-\frac18XC_{21}+\frac18X^2C_{20}.
\label{doubexp}
\eea
We have checked all these non-zero coefficients, and moreover we have verified 
that the double poles for the remaining invariants whose coefficients we are
computing vanish as they should. Of course the double pole contributions 
can in principle come from non-planar as well as planar diagrams. However,
one can check that the only double-pole contribution from a non-planar
diagram to one of the group structures whose divergent contribution
we have computed
is that from the diagram Fig.~\ref{nonplanar} (which contributes to 
$\alpha_{22}$). Indeed, diagrams with three-point
gauge vertices only have simple poles and the majority of non-planar diagrams
are of this type. Including this double-pole contribution along with 
those from the planar diagrams in Table.~\ref{restabl}, we find that the 
double pole proportional
to $X_2$ in Eq.~(\ref{gstruct}) is indeed cancelled. Of course the double poles
corresponding to the invariants with coefficients
$\alpha_{14,15}$, $\alpha_{18}$, $\alpha_{20}$, $\alpha_{26}$,
$\alpha_{29}$ and $\alpha_{30-33}$ should also cancel, but this we have not 
checked.

We note that the diagrams listed in Table~\ref{restabk}
 consist of insertions of a
two-loop contribution to the gauge two-point function. These would be relevant
to a superspace calculation of the corrections to the Chern-Simons level 
$k$; also required would be the similar contributions to the ghost two-point
function; and the two-loop corrections to the $V\Phibar\Phi$ vertex. Such 
calculations have been performed in components
\cite{semwu}, but there may be some interest 
in corroborating them in the superspace context.

Our final result for the four-loop anomalous dimension is
\bea
(8\pi)^4\gamma_{\Phi}^{(4)}&=&\frac23Y_3+\frac{\pi^2}{4}Y_4
-\frac43Z_1+\left(4-\frac23\pi^2\right)(4W_1-Z_2)\nn
&+&\left(8-\frac53\pi^2\right)W_2-\frac13\pi^2W_3
+\frac23\pi^2W_4\nn
&+&\left[2\left(1-\frac18\pi^2\right) X
-\left(1-\frac14\pi^2\right)C_G\right]\left(\frac13U_1-U_2\right)\nn
&-&4\left(6+\pi^2\right)C_{40}+\left(32+\frac{17}{6}\pi^2\right)C_{31}
-\frac12\left(25+\frac{23}{24}\pi^2\right)C_{22}\nn
&+&\alpha_{14}C_{13}+
\alpha_{15}F_4
+X\left[-(8+3\pi^2)C_{30}+\left(2+\frac{19}{6}\pi^2\right)C_{21}
+\alpha_{18} C_{12}\right]\nn
&+&X^2[-(2+\pi^2)C_{20}+\alpha_{20}C_{11}]
-\frac18\pi^2X^3C_R+\left(16-\frac73\pi^2\right)X_2\nn
&-&\frac23X_4
+\left[(8-3\pi^2)X-2\pi^2C_R+
\alpha_{26}C_G\right]X_1+\frac{16\pi^2}{3}X_5C_R\nn
&+&
(-\pi^2X+\alpha_{29}C_G)X_5+\alpha_{30}X_3
+\alpha_{31}\tr(C_R\{R_A,R_B\}R_C)R_AR_BR_C\nn
&+&\alpha_{32}X_6
+\alpha_{33}d_{CDA}d_{CDB}R_AR_B.
\label{finalres}
\eea

As we explained in the introduction, we believe that the remaining 
undetermined coefficients may be determined by comparison with 
a small number of the known superconformal
theories.

\begin{table}
\begin{center}
\begin{tabular}{|c| c c c c c|}\hline
&$C_{40}$&$C_{31}$&$C_{22}$&$C_{13}$&$F_4$\\ \hline
$S_1$&1&$-\frac12$&$\frac{1}{16}$&0&0\\ [0.5ex]
$S_2$&1&$-\frac32$&$\frac{41}{48}$&$-\frac{33}{192}$&$\frac{5}{24}$\\ [0.5ex] 
$S_3$&1&$-\frac34$&$\frac{1}{8}$&0&0\\ [0.5ex]
$S_4$&1&$-\frac54$&$\frac{13}{24}$&$-\frac{1}{12}$&0\\ [0.5ex]
$S_5$&1&$-\frac32$&$\frac56$&$-\frac{47}{288}$&$\frac{7}{36}$\\ [0.5ex]
$S_6$&1&$-2$&$\frac{65}{48}$&$-\frac{29}{96}$&$\frac{5}{12}$ \\ [0.5ex]
$S_7$&1&$-\frac52$&$\frac{33}{16}$&$-\frac{17}{32}$&$1$ \\ [0.5ex]
$S_8$&1&$-1$&$\frac{3}{8}$&$-\frac{7}{128}$&$\frac{1}{16}$ \\ [0.5ex]
$S_9$&1&$-\frac32$&$\frac{13}{16}$&$-\frac{5}{32}$&$\frac{1}{4}$ \\ [0.5ex]
$S_{10}$&1&$-\frac54$&$\frac{9}{16}$&$-\frac{3}{32}$&$\frac{1}{8}$ \\ [0.5ex]
$S_{11}$&1&$-\frac74$&$\frac{33}{32}$&$-\frac{13}{64}$&$\frac{1}{4}$ \\ [0.5ex]
$S_{12}$&1&$-\frac32$&$\frac{3}{4}$&$-\frac18$&$0$ \\ [0.5ex]
$S_{13}$&1&$-\frac32$&$\frac{3}{4}$&$-\frac{1}{8}$&$\frac{1}{16}$ \\ [0.5ex]
$S_{14}$&1&$-\frac74$&$1$&$-\frac{3}{16}$&$\frac{1}{8}$ \\ [0.5ex]
$S_{15}$&1&$-\frac{21}{12}$&$\frac{103}{96}$&$-\frac{43}{192}$&$\frac{7}{24}$
 \\ [0.5ex]
$S_{16}$&1&$-2$&$\frac{21}{16}$&$-\frac{9}{32}$&$\frac{1}{4}$ \\ [0.5ex]
$S_{17}$&1&$-\frac94$&$\frac{27}{16}$&$-\frac{13}{32}$&$\frac{5}{8}$ \\ [0.5ex]
$S_{18}$&0&$1$&$-\frac{11}{8}$&$\frac{7}{16}$&$-1$ \\ [0.5ex]
$S_{19}$&0&$1$&$-\frac{7}{8}$&$\frac{3}{16}$&$0$ \\ [0.5ex]
$S_{20}$&0&$0$&$1$&$-\frac{3}{8}$&2 \\ [0.5ex]
$S_{21}$&0&$0$&$\frac32$&$-\frac{5}{8}$&$1$ \\ [0.5ex]
$S_{22}$&0&$0$&$2$&$-1$&$2$ \\ [0.5ex]
$S_{23}$&0&$0$&$\frac12$&$-\frac14$&$2$ \\ [0.5ex]
$S_{24}$&0&$0$&$1$&$-\frac{7}{18}$&$\frac43$ 
\\ \hline

\end{tabular}
\caption{Decompositions into group invariants for diagrams of 
type Fig.~\ref{fig3}(a)}
\label{group}
\end{center}
\end{table}

\begin{table}
\begin{center}
\begin{tabular}{|c| c c c |}\hline
&$C_{30}$&$C_{21}$&$C_{12}$\\ \hline
$S_{X1}$&1&$-\frac34$&$\frac{1}{6}$\\ [0.5ex]
$S_{X2}$&1&$-\frac{7}{12}$&$\frac{5}{48}$\\ [0.5ex] 
$S_{X3}$&1&$-\frac34$&$\frac{5}{32}$\\ [0.5ex]
$S_{X4}$&1&$-1$&$\frac14$\\ [0.5ex]
$S_{X5}$&1&$-\frac14$&$0$\\ [0.5ex]
$S_{X6}$&0&$1$&$-\frac38$
\\ \hline

\end{tabular}
\caption{Decompositions into group invariants for diagrams of 
type Fig.~\ref{fig3}(b)}
\label{groupa}
\end{center}
\end{table}

\begin{table}
\begin{center}
\begin{tabular}{|c| c c c c c c c|}\hline
&&$I_4$&$I_{22}$&$I_{4bbb}$&$I_{42bbc}$&$I_{422qAbBd}$&\\ [0.5ex]
$Y_3$&$\frac{1}{12}$&1&0&0&0&0&$Y_3$\\ [0.5ex]
$Y_4$&$\frac{1}{8}$&0&0&1&0&0&$Y_4$\\ [0.5ex]
$[A][B][](AB)$&$-1$&$-\frac74$&$-\frac58$&1&1&$-2$&$W_1$\\ [0.5ex]
$[AB][](AB)$&$-1$&$-2$&0&1&0&0&$W_2-\frac{1}{12}C_GU$\\ [0.5ex]
$[(AB)][](AB)$&$\frac12$&$-2$&0&0&0&0&$W_2-\frac{1}{12}C_GU_1$\\ [0.5ex]
$[(AB)(AB)][][]$&$\frac14$&$-2$&0&0&0&0&$Z_2-\frac{1}{4}C_GU_2$\\ [0.5ex]
$[(AB)][(AB)][]$&
$\frac14$&0&0&$-2$&0&0&$W_3-\frac14C_GU_2+\frac{1}{12}C_GU_1$\\ [0.5ex]
$[A][(AB)][B][]$&$-\frac12$&$\frac14$&$-\frac58$&$-1$&1&$-2$&
$\frac12W_2-W_3+W_4+\frac12Z_2$\\ [0.5ex]
$[(AB)][AB][]$&
$-1$&$\frac14$&$-\frac58$&$-1$&1&$-2$&
$W_3-\frac14C_GU_2+\frac{1}{12}C_GU_1$\\ [0.5ex]
$[ABAB][][]$&$\frac12$&$0$&0&1&0&0&$Z_2-\frac{1}{2}C_GU_2$\\ [0.5ex]
$[ABA][B][]$&$1$&$-\frac14$&$\frac58$&$1$&$-1$&$2$&$-\frac12Z_2+\frac{1}{4}C_GU_2
-\frac{1}{12}C_GU_1$\\ [0.5ex]
$(AB)[][][](AB)$&$\frac{1}{12}$&$-2$&0&0&0&0&$Z_1-\frac{1}{4}C_GU_1$\\ [0.5ex]
$[AA][][]X_{AA}$&
$\frac{1}{2}$&$1$&0&$-\frac12$&0&0&$X\left(U_1-\frac12U_2\right)$
\\ [0.5ex]
$[A][A]X_{AA}$&$\frac{1}{2}$&$0$&$\frac12$&0&0&0&$XU_2$\\ [0.5ex]
$[(AA)][][]X_{AA}$&$-\frac{1}{2}$&$0$&$\frac12$&0&0&0&$XU_2$\\ \hline
\end{tabular}
\caption{Results for diagrams of type Fig.~\ref{fig3}(g)}
\label{restaba}
\end{center}
\end{table}

\bigskip
\begin{table}
\begin{center}
\begin{tabular}{|c| c c c c c c c|}\hline
&&$I_4$&$I_{22}$&$I_{4bbb}$&$I_{42bbc}$&$I_{422qAbBd}$&\\ [0.5ex]
$(44)(33)(22)(11)$&$\frac14$&4&0&0&0&0&$S_1$\\ [0.5ex]
$(233)(13)(112)$&$-\frac12$&0&0&4&0&0&$S_5$\\ [0.5ex]
$(66)4523(11)$&$\frac12$&0&0&$-2$&0&0&$S_3$\\ [0.5ex]
$(355)412(11)$&$-1$&0&0&$-2$&0&0&$S_4$\\ [0.5ex]
$(24)(13)(24)(13)$&$1$&0&0&$-2$&0&0&$S_8$\\ [0.5ex]
$(36)516(14)$&$1$&0&0&0&$-2$&0&$S_9$\\ [0.5ex]
$(25)(14)52(13)$&$-2$&0&0&$-1$&0&0&$S_{10}$\\ [0.5ex]
$(35)4(15)2(13)$&$-1$&0&0&0&$-2$&0&$S_{11}$\\ [0.5ex]
$(46)6513(12)$&$2$&$-\frac14$&$\frac58$&0&$-1$&2&$S_{12}$\\ [0.5ex]
$(34)(34)(12)(12)$&$1$&0&0&$-1$&2&0&$S_{13}$\\ [0.5ex]
$(35)(45)12(12)$&$-2$&0&0&$-1$&2&0&$S_{14}$\\ [0.5ex]
$(223)(113)(12)$&$-1$&4&0&0&0&0&$S_5$\\ [0.5ex]
$(234)(14)1(12)$&$2$&0&0&$1$&0&0&$S_{15}$\\ [0.5ex]
$(334)4(11)(12)$&$1$&4&0&$-2$&0&0&$S_6$\\ [0.5ex]
$(345)511(12)$&$-2$&$\frac14$&$-\frac58$&0&$1$&$-2$&$S_6$\\ [0.5ex]
$(2233)(11)(11)$&$-\frac12$&4&0&0&0&0&$S_2$\\ [0.5ex]
$(33)(44)(11)(22)$&$\frac14$&0&4&0&0&0&$S_7$\\ [0.5ex]
$(22)(1133)(22)$&$-\frac14$&0&4&0&0&0&$S_2$\\ [0.5ex]
$(33)4(114)(23)$&$1$&0&2&0&0&0&$S_6$\\ [0.5ex]
$(34)51(15)(24)$&$-2$&$-2$&$1$&$1$&0&0&$S_{14}$\\ [0.5ex]
$(23)(14)(14)(23)$&$1$&$-2$&$1$&0&0&0&$S_{13}$\\ [0.5ex]
$(35)6161(24)$&$1$&$-\frac52$&$\frac94$&$2$&$-2$&$4$&$S_{16}$\\ [0.5ex]
$(34)5(15)1(23)$&$-1$&$-2$&$1$&2&0&0&$S_{17}$\\ [0.5ex]
$(34)(55)11(22)$&$-1$&0&2&0&0&0&$S_7$\\ [0.5ex]
$(45)6611(23)$&$1$&$-2$&$1$&2&0&0&$S_7$\\ [0.5ex]
$(3\alpha)(4\alpha)1(2\alpha)$&
$1$&$-\frac34$&$-\frac18$&0&$1$&$-2$&$S_{18}$\\ [0.5ex]
$(5\alpha)4\alpha2(1\alpha)$&
$-1$&$-\frac18$&$\frac{5}{16}$&1&$-1$&$1$&$S_{19}$\\ [0.5ex]
\hline

\end{tabular}
\caption{Results for diagrams of type Fig.~\ref{fig3}(a),(f)}
\label{restabb}
\end{center}
\end{table}

\bigskip
\begin{table}
\begin{center}
\begin{tabular}{|c| c c c c c c c|}\hline
&&$I_4$&$I_{22}$&$I_{4bbb}$&$I_{42bbc}$&$I_{422qAbBd}$&\\ [0.5ex]
$(\alpha\alpha)(3\alpha)(2\alpha)$&$\frac{1}{24}$&$1$&
$\frac12$&0&$0$&$0$&$S_{20}$\\ [0.5ex]
$(3\alpha\alpha)\alpha(1\alpha)$&$\frac{1}{72}$&$0$&$0$&$-\frac12$&
$0$&$0$&$S_{21}$\\ [0.5ex]
$(3\alpha)(\alpha\alpha)(1\alpha)$&
$-\frac{1}{192}$&$0$&$0$&$-4$&$-4$&$0$&$S_{20}$\\ [0.5ex]
$(3\alpha)\alpha\alpha(1\alpha)$&
$\frac{1}{24}$&$\frac12$&$-\frac{5}{4}$&$-1$&$3$&$-4$&
$S_{20}$\\ [0.5ex]
$(3\alpha)\alpha\alpha(1\alpha)'$&
$\frac{1}{8}$&$-\frac14$&$\frac58$&$0$&$-1$&$2$&$F_4$\\ [0.5ex]
$(ab)(ab)(ab)$&$-\frac{1}{4}$&$\frac58$&$-\frac{9}{16}$&$-1$&$\frac12$&$-1$&
$S_{22}$\\ [0.5ex]
$(\alpha\alpha)4\alpha(2\alpha)$
&$-\frac{1}{12}$&$0$&$0$&$\frac12$&$0$&$0$&$S_{23}$\\ [0.5ex]
$(3\alpha\alpha)(1\alpha\alpha)$
&$\frac{1}{48}$&$0$&$0$&$1$&$0$&$0$&$S_{24}$\\ [0.5ex]
Fig.~\ref{fig3a}&$-\frac{1}{8}$&$-1$&$\frac12$&$0$&$0$&$0$&$S_{20}$\\ 
\hline

\end{tabular}
\caption{Results for graphs with 4-point gauge vertex and graphs of 
type Fig.~\ref{fig3}(c),(d)}
\label{restabc}
\end{center}
\end{table}

\bigskip
\begin{table}
\begin{center}
\begin{tabular}{|c| c c c c c c c|}\hline
&&$I_4$&$I_{22}$&$I_{4bbb}$&$I_{42bbc}$&$I_{422qAbBd}$&\\ [0.5ex]
$(3A)C(1B)\{132\}$&1&0&0&0&$-1$&0&$X_5C_R+\frac14T_R(C_{21}-\frac38C_{12})$
\\ [0.5ex]
$(3A)B(1C)\{123\}$&1&$\frac12$&$-\frac54$&$-1$&$2$&$-4$&$X_5C_R-\frac14
T_R(C_{21}-\frac38C_{12})$\\[0.5ex]
$(3A')C'(1B')\{132\}$&$\frac{1}{16}$&0&0&0&$-1$&0&$C_{22}-\frac38C_{13}$
\\ [0.5ex]
$(3A')B'(1C')\{123\}$&$-\frac{1}{16}$&$\frac12$&$-\frac54$&$-1$&$2$&$-4$&
$C_{22}-\frac38C_{13}$\\[0.5ex]
$(A3)A(B1)\{(12)3\}$&$-2$&0&0&$-\frac12$&0&0&$X_5C_R$\\[0.5ex]
$(AB2)(C1)\{112\}$&$-2$&$-2$&0&0&0&0&$X_5\left(C_R-\frac13C_G\right)$\\[0.5ex]
$(AB2)(B1)\{1(12)\}$&$2$&$-2$&0&0&0&0&$X_5\left(C_R-\frac13C_G\right)$\\[0.5ex]
$(AA2)(B1)\{(11)2\}$&$1$&$0$&0&$-2$&0&0&$X_5\left(C_R-\frac13C_G\right)$
\\[0.5ex]
$(3A)B(1A)\{(13)2\}$&$-1$&0&0&0&$-1$&0&$X_5(C_R-\frac38C_G)$\\ [0.5ex]
\hline

\end{tabular}
\caption{Results for graphs contributing to $X_5C_R$, and similar topologies}
\label{restabd}
\end{center}
\end{table}

\begin{table}
\begin{center}
\begin{tabular}{|c| c c c c c c|}\hline
&&$I_4$&$I_{22}$&$I_{4bbb}$&$I_{42bbc}$&\\ [0.5ex]
$(222)(111)X_{12}$&$-1$&$1$&0&$0$&$0$&$XS_{X1}$\\ [0.5ex]
$(1122)(11)X_{11}$&$-1$&$0$&1&$0$&$0$&$XS_{X2}$\\ [0.5ex]
$(33)(22)(11)X_{22}$&$\frac12$&$0$&$1$&$0$&$0$&$XS_{X5}$\\ [0.5ex]
$(233)1(11)X_{12}$&$2$&$0$&$\frac12$&$0$&$0$&$XS_{X1}$\\ [0.5ex]
$(44)32(11)X_{23}$&$-1$&$0$&$\frac12$&$0$&$0$&$XS_{X5}$\\ [0.5ex]
$(2a)(1a)X_{aa}$&$\frac16$&$0$&$1$&$0$&$0$&$XC_{21}$\\ [0.5ex]
$(23)(13)(12)X_{12}$&$-4$&$\frac12$&$-\frac14$&$0$&$0$&$XS_{X3}$\\ [0.5ex]
$(23)(13)(12)X_{13}$&$-1$&$0$&$0$&$1$&$0$&$XS_{X3}$\\ [0.5ex]
$(34)41(12)X_{14}$&$1$&$0$&$0$&$\frac12$&$0$&$XS_{X4}$\\ [0.5ex]
$(34)41(12)X_{13}$&$2$&$1$&$-\frac12$&$-\frac12$&$0$&$XS_{X4}$\\ [0.5ex]
$(\alpha3)\alpha(\alpha1)X_{\alpha2}$
&$-\frac12$&$-1$&$\frac12$&$\frac12$&$-\frac12$&$XS_{X6}$
\\ \hline

\end{tabular}
\caption{Results for diagrams of type Fig.~\ref{fig3}(b)}
\label{restabe}
\end{center}
\end{table}

\bigskip

\begin{table}
\begin{center}
\begin{tabular}{|c| c c c c c c c c|}\hline
&&$I_4$&$I_{22}$&$I_{4bbb}$&$I_{42bbc}$&$J_4$&$J_5$&\\ [0.5ex]
$(AB)\{(1B)(1A)\}X_{AB}$&$-\frac12$&$2$&0&$0$&$0$&2&2&
$X(X_1-\frac14C_GT_RC_R)$
\\ [0.5ex]
$(AB)\{(1B)(1A)\}X_{1A}$&$-1$&$2$&0&$0$&0&0&$0$&$X(X_1
-\frac14C_GT_RC_R)$
\\ [0.5ex]
$(AC)\{(1B)A1\}X_{AB}$&$1$&$0$&$1$&$0$&$0$&1&1&$X(X_1
-\frac14C_GT_RC_R)$
\\ [0.5ex]
$(AB)\{(1C)1A\}X_{AC}$&$1$&$0$&$1$&$0$&$0$&1&1&$X(X_1
-\frac14C_GT_RC_R)$
\\ [0.5ex]
$(AD)\{1CB1\}X_{BC}$&$-1$&$0$&$1$&$0$&$0$&1&1&$X(X_1
-\frac12C_GT_RC_R)$
\\ [0.5ex]
$(AC)\{1D1B\}X_{BD}$&$
-\frac12$&$-2$&$2$&$1$&$-2$&0&0&$X(X_1
-\frac12C_GT_RC_R)$
\\ [0.5ex]
$(AC)\{1D1B\}X_{1A}$&$-1$&$-2$&$0$&$1$&$0$&0&0&$X(X_1
-\frac14C_GT_RC_R)$
\\ [0.5ex]
$(AC)\{1(BB)1\}X_{BB}$&$1$&$0$&$2$&$0$&$0$&0&0&$XX_1$\\ [0.5ex]
$(AB)\{(1AA)1\}X_{AA}$&$1$&$0$&$-2$&$0$&$0$&0&0&$X(X_1
-\frac12C_GT_RC_R)$\\ 
\hline

\end{tabular}
\caption{Results for diagrams contributing to $XX_1$}
\label{restabf}
\end{center}
\end{table}

\bigskip

\begin{table}
\begin{center}
\begin{tabular}{|c| c c c c c|}\hline
&&$I_4$&$J_4$&$I_{42bbc}$&\\ [0.5ex]
Fig.~\ref{fig4}(a)&$\frac{1}{12}$&$2$&$-2$&$-2$&$X_4C_R$\\ [0.5ex]
Fig.~\ref{fig4}(b)&$-\frac{1}{6}$&$1$&$-1$&$0$&$X_4C_R$\\ 
\hline

\end{tabular}
\caption{Results for diagrams contributing to $X_4C_R$}
\label{restabg}
\end{center}
\end{table}

\bigskip

\begin{table}
\begin{center}
\begin{tabular}{|c| c c c c c|}\hline
&&$I_4$&$I_{22}$&$I_{4bbb}$&\\ [0.5ex]
$(11)(11)X_{11}X_{11}$&
$-\frac18$&$0$&$1$&$0$&$X^2(C_{20}-\frac16C_{11})$
\\ [0.5ex]
$(22)(11)X_{12}X_{12}$&
$\frac12$&$-\frac12$&$\frac14$&$0$&$X^2(C_{20}-\frac14C_{11})$
\\ [0.5ex]
$(22)(11)(X^2)_{12}$&$1$&$0$&$0$&$-\frac12$&$X^2(C_{20}-\frac14C_{11})$\\[0.5ex]
$(11)(X^3)_{11}$&$-\frac12$&$0$&$0$&$\frac18$&$X^3C_R$\\
\hline

\end{tabular}
\caption{Results for diagrams contributing to $X^2C_R^2$ and $X^3C_R$}
\label{restabi}
\end{center}
\end{table}
\bigskip

\begin{table}
\begin{center}
\begin{tabular}{|c| c c c c c|}\hline
&&$I_4$&$I_{22}$&$I_{4bbb}$&\\ [0.5ex]
$(ABC)X_{1A}$&$1$&$1$&$0$&0&$XX_5$\\ [0.5ex]
$(ABB)X_{1B}$&$-1$&$1$&$0$&0&$XX_5$\\ [0.5ex]
$(ABB)X_{1A}$&$-\frac12$&$0$&0&$1$&$XX_5$\\ 
\hline

\end{tabular}
\caption{Results for diagrams contributing to $XX_5$}
\label{restabj}
\end{center}
\end{table}
\bigskip

\begin{table}
\begin{center}
\begin{tabular}{|c| c c c c |}\hline
&&$I_4$&$I_{4bbb}$&\\ [0.5ex]
$(2a)(1b)\{(1b)(2a)\}$&$\frac{1}{24}$&$1$&0&$C_{22}-\frac14C_{13}$
\\ [0.5ex]
$(2A')(1B')\{(1B')(2A')\}$&$\frac{1}{6}$&$1$&0&$C_{22}-\frac14C_{13}$
\\ [0.5ex]
$(2A')(1C')\{1D'2B'\}$&$-\frac{1}{8}$&$1$&$-\frac12$&$C_{22}-\frac14C_{13}$
\\ [0.5ex]
$(2A')(1\alpha)\{1\alpha\alpha\}$&$-\frac{1}{8}$&$0$&1&$C_{22}-\frac14C_{13}$
\\ [0.5ex]
$(2A)(1B)\{(1B)(2A)\}$&$-2$&$1$&0&$(X_1-\frac14C_GT_RC_R)(C_R-\frac14C_G)$\\ 
[0.5ex]
$(2A)(1C)\{1D2B\}$&$2$&$1$&$-\frac12$&$(X_1-\frac14C_GT_RC_R)(C_R-\frac14C_G)$
\\ [0.5ex]
$(2A)(1\alpha)\{1\alpha\alpha\}$&$1$&$0$&$2$&$\frac14C_GT_R$\\
\hline
\end{tabular}
\caption{Results for two-loop vector two-point insertion diagrams}
\label{restabk}
\end{center}
\end{table}
\bigskip

\begin{table}
\begin{center}
\begin{tabular}{|c| c c c c c c c|}\hline
&&$I_4$&$I_{22}$&$I_{4bbb}$&$I_{42bbc}$&$I_{422qAbBd}$&\\ [0.5ex]
$(AB)(CD)$&$1$&$-4$&0&$1$&$1$&0&$X_2-\frac{1}{16}T_RC_{12}
+\frac12C_GX_5$\\ [0.5ex]
$(AC)(BC)$&$-1$&$-4$&0&$1$&$1$&0&$X_2-\frac{3}{32}T_RC_{12}
+\frac58C_GX_5-\frac14iX_6$\\ [0.5ex]
$(AB)(BC)$&$-1$&$-\frac{15}{4}$&$-\frac{5}{8}$&$1$&$2$&$-2$&$X_2
-\frac{3}{32}T_RC_{12}
+\frac58C_GX_5-\frac14iX_6$\\ [0.5ex]
$(AB)(AB)$&$\frac12$&$-4$&0&$1$&$2$&0&$X_2-\frac{3}{32}T_RC_{12}
+\frac58C_GX_5-\frac14iX_6$\\ [0.5ex]
$(AA)(BC)$&$-1$&$-2$&0&$0$&$0$&0&$X_2-\frac{1}{16}T_RC_{12}
+\frac12C_GX_5$\\ [0.5ex]
$(AA)(AB)$&$1$&$-2$&$0$&$0$&$0$&0&$X_2-\frac{3}{32}T_RC_{12}
+\frac58C_GX_5-\frac14iX_6$\\ [0.5ex]
$(ABCD)$&$-\frac14$&$\frac14$&$-\frac{5}{8}$&$-1$&$1$&$-2$&$X_2
-\frac{3}{32}T_RC_{12}
+\frac58C_GX_5-\frac14iX_6$\\ 
$(AABB)$&$-\frac18$&$0$&0&$4$&$0$&0&$X_2-\frac{3}{32}T_RC_{12}
+\frac58C_GX_5-\frac14iX_6$
\\ \hline

\end{tabular}
\caption{Results for diagrams contributing to $X_2$}
\label{restabl}
\end{center}
\end{table}

\bigskip

\section{Conclusions}
As we stated in the introduction, it should be possible to exploit our results 
in Eq.~(\ref{finalres})
in order to verify the superconformality properties of various superconformal
theories, as given explicitly either in terms of a ``quiver'' description 
or using 3-algebra structures. 
Our results (presented for the case of an ordinary simple gauge group) may 
require some adaptation in order to cast them in
one or other of these forms; but we believe that this should be possible by
inspection of the diagrammatic structure, in view of the fact that we have 
made no assumptions beyond standard group commutation relations in reducing our
results to a standard basis of invariants. The ``quiver'' description  
is based on $U(N)$ (rather than $SU(N)$), and therefore identities such as 
those in Eq.~(\ref{groupid})
which we have used in our reduction to a basis of invariants, would require
some modification, as detailed for instance in Ref.~\cite{bonora}; but these
changes would have no effect at leading $N$, to which we are restricted anyway
owing to our focus on planar diagrams.
Given the large number of available
candidates for superconformal theories, we expect that specialisation of our 
results to these various cases will provide compelling evidence for the
superconformality at this highly non-trivial order.

One possible subtlety is the fact that renormalisation-group
quantities such as the anomalous
dimension are scheme-dependent beyond lowest order. This means that in 
principle one might have to adjust the couplings order by order so as to 
achieve finiteness\cite{DRTJ,EKT}. 
This could be accommodated by parametrising all possible 
redefinitions of the couplings. It is perhaps worth recalling that in 
four dimensions, the finiteness properties of $\Ncal=4$ and $\Ncal=2$ 
supersymmetric theories are manifest to all orders in the $\Ncal =1$
superfield description once the field content has been specified (assuming 
a supersymmetric regulator such as DRED).
However, when finite $\Ncal =1$ theories are constructed, the finiteness
is obtained through an order-by-order adjustment of the couplings. On the whole
it seems to us likely that the superconformal situation in three dimensions 
will resemble the $\Ncal =4$ and $\Ncal=2$ situation in four 
dimensions in cases where exact, explicit formulations can be presented for 
the action, such as the BLG and ABJ/ABJM cases. However, in cases such as the 
superconformal theories conjectured to 
exist in Ref.~\cite{GT} through a process of interpolation, and given explicitly
only to lowest order in Ref.~\cite{penatib}; and indeed the whole range of 
superconformal theories found in Refs.~\cite{penatia}, \cite{penatib} by 
solving the lowest order conformal invariance conditions,
order-by-order redefinitions of the couplings will be needed. The coupling 
redefinitions should merely result in a slight reduction 
of the large redundancy in the system of equations rising from requiring
superconformality of the various candidate theories; and so the 
superconformality check will still be compelling.    
  
Of course further weight would be given to any superconformality checks by 
continuing the computation of the remaining unknown coefficients in 
Eq.~(\ref{finalres}). This would be hugely simplified if we could understand 
in advance which diagrams with 3-point gauge vertices will yield a vanishing 
contribution. In any case the remainder of the computation is certainly not 
insuperable, merely somewhat laborious. The extension to the non-planar 
diagrams is also in principle feasible, though we do not at present have 
available a convenient basis of momentum integrals already tabulated for this 
case. On the other hand, many of the non-planar diagrams may not actually
contribute since most of them will contain 3-point gauge vertices. 

\bigskip
\bigskip

\noindent
{\large{\bf Acknowledgements\\}}
One of us (CL) was supported by a University of Liverpool studentship. IJ is 
grateful for useful discussions with Tim Jones.

\bigskip
\bigskip

\noindent

{\large{\bf{Appendix}\hfil}}

In this appendix we list our superspace and supersymmetry conventions.
We use a metric signature $(+--)$ so that a possible choice of $\gamma$ matrices
is $\gamma^0=\sigma_2$, $\gamma^1=i\sigma_3$, $\gamma^2=i\sigma_1$
with
\be
(\gamma^{\mu})_{\alpha}{}^{\beta}=(\sigma_2)_{\alpha}{}^{\beta}, 
\ee
etc. We then have 
\be
\gamma^{\mu}\gamma^{\nu}=\eta^{\mu\nu}-i\epsilon^{\mu\nu\rho}\gamma_{\rho}.
\ee
We have\cite{penatib}
 two complex two-spinors $\theta^{\alpha}$ and $\theta^{\alpha}$ 
with indices raised and lowered according to
\be
\theta^{\alpha}=C^{\alpha\beta}\theta_{\beta},\quad
\theta_{\alpha}=\theta^{\beta}C_{\beta\alpha},
\ee
with $C^{12}=-C_{12}=i$. We then have
\be
\theta_{\alpha}\theta_{\beta}=C_{\beta\alpha}\theta^2,\quad
\theta^{\alpha}\theta^{\beta}=C^{\beta\alpha}\theta^2,
\ee
where 
\be
\theta^2=\frac12\theta^{\alpha}\theta_{\alpha}.
\ee
The supercovariant derivatives are defined by
\bea
D_{\alpha}=&\pa_{\alpha}+\frac{i}{2}\thetabar^{\beta}\pa_{\alpha\beta},\\
\Dbar_{\alpha}=&\pabar_{\alpha}+\frac{i}{2}\theta^{\beta}\pa_{\alpha\beta},
\eea
where
\be
\pa_{\alpha\beta}=\pa_{\mu}(\gamma^{\mu})_{\alpha\beta},
\label{pdef}
\ee
satisfying
\be
\{D_{\alpha},\Dbar_{\beta}\}=i\pa_{\alpha\beta}.
\ee
We also define
\be
d^2\theta=\frac12d\theta^{\alpha}d\theta_{\alpha}
\quad d^2\thetabar=\frac12d\thetabar^{\alpha}d\thetabar_{\alpha},
d^4\theta=d^2\theta d^2\thetabar,
\ee
so that
\be
\int d^2\theta\theta^2=\int d^2\thetabar\thetabar^2=-1.
\ee
The vector superfield $V(x,\theta,\thetabar)$ is expanded in Wess-Zumino gauge
as
\be
V=i\theta^{\alpha}\thetabar_{\alpha}\sigma+\theta^{\alpha}\thetabar^{\beta}
A_{\alpha\beta}-\theta^2\thetabar^{\alpha}\lambdabar_{\alpha}
-\thetabar^{2}\theta^{\alpha}\lambda_{\alpha}+\theta^2\thetabar^{2}D,
\ee
and the chiral field is expanded as
\be
\Phi=\phi(y)+\theta^{\alpha}\psi_{\alpha}(y)-\theta^2F(y),
\ee
where
\be
y^{\mu}=x^{\mu}+i\theta\gamma^{\mu}\thetabar.
\ee

Here is the list of results for the divergences of our basis of momentum
integrals\cite{MSS,MSSa}:
\bea
I_4&=&
\frac{1}{(8\pi)^4}\left(-\frac{1}{2\epsilon^2}+\frac{2}{\epsilon}\right)\\
I_{22}&=&
-\frac{1}{(8\pi)^4\epsilon^2}\\
I_{4bbb}&=&
\frac{1}{(8\pi)^4}\frac{\pi^2}{2\epsilon}\\
I_{42bbc}&=&
\frac{1}{(8\pi)^4}\frac{2}{\epsilon}\\
I_{422qAbBd}&=&\frac{1}{(8\pi)^4}
\left[\frac{1}{4\epsilon^2}+\frac{1}{\epsilon}
\left(\frac54-\frac{\pi^2}{12}\right)\right]
\eea

The table of group structures may be readily obtained using the
following easily derived but useful group identities:
\bea
R_BR_AR_B&=&\left(C_R-\frac12C_G\right)R_A,\nn
f^{ABE}f^{CDE}R_AR_CR_BR_D&=&0,\nn
R_AR_BR_CR_AR_BR_C&=&C_{30}-\frac32C_{21}+\frac12C_{12},\nn
R_AR_BR_CR_DR_AR_BR_CR_D&=&C_{40}-3C_{31}+\frac{11}{4}C_{22}-\frac34C_{13}
+F_4,\nn
R_AR_BR_CR_AR_DR_BR_CR_D&=&C_{40}-\frac52C_{31}+2C_{22}-\frac12C_{13}
+F_4,\nn
R_AR_BR_CR_DR_AR_CR_BR_D&=&C_{40}-2C_{31}+\frac32C_{22}-\frac38C_{13}
+F_4,\nn
f^{ABC}R_AR_DR_CR_ER_DR_BR_E&=&-i\left(\frac12C_{31}-\frac34C_{22}+
\frac14C_{13}-F_4\right),\nn
f^{ABF}f^{CDF}R_AR_CR_ER_DR_BR_E&=&\frac14C_{22}-\frac18C_{13}+F_4,\nn
f^{ABF}f^{CDF}R_AR_CR_ER_BR_DR_E&=&F_4,\nn
f^{ABC}f^{DEF}R_AR_DR_BR_ER_CR_F&=&-\frac14C_{22}+\frac18C_{13}-F_4
\label{groupid}
\eea

\end{document}